\documentclass[12pt,a4paper]{article}
\makeatletter
\def\eqnarray{%
\stepcounter{equation}%
\let\@currentlabel=\theequation
\global\@eqnswtrue
\global\@eqcnt\z@
\tabskip\@centering
\let\\=\@eqncr
$$\halign to \displaywidth\bgroup\@eqnsel\hskip\@centering
$\displaystyle\tabskip\z@{##}$&\global\@eqcnt\@ne
\hfil$\displaystyle{{}##{}}$\hfil
&\global\@eqcnt\tw@$\displaystyle\tabskip\z@{##}$\hfil
\tabskip\@centering&\llap{##}\tabskip\z@\cr}
\makeatother
\newcommand{\ket}[1]{{\vert{#1}\rangle}}

\newcommand{\fukuso}{{\mathbf C}}

\begin{document}

\title{\sl A Generalization of the Preceding Paper 
``A Rabi Oscillation in Four and Five Level Systems"}
\author{
  Kazuyuki FUJII 
  \thanks{E-mail address : fujii@yokohama-cu.ac.jp }\\
  Department of Mathematical Sciences\\
  Yokohama City University\\
  Yokohama, 236--0027\\
  Japan
  }
\date{}
\maketitle
%\thispagestyle{empty}
%
%
%  gaiyou
%
%
\begin{abstract}
  In the preceding paper quant-ph/0312060 we considered a general model of 
  an atom with n energy levels interacting with n--1 external laser fields 
  and constructed a Rabi oscillation in the case of n =3, 4 and 5.
  
  In the paper we present a systematic method getting along with computer 
  to construct a Rabi oscillation in the general case. 
\end{abstract}
%

%\newpage

%
%
%     Honbun
%
%

\newpage

In the preceding paper \cite{FHKW2} (and \cite{KF4}) we considered a general 
model of an atom with n energy levels interacting with n--1 external laser 
fields and constructed a Rabi oscillation in the case of n = 3, 4 and 5. 
Concerning more realistic model on two level system see \cite{FHKW3} and 
\cite{FHKW4}.

The purpose of this paper is to present a systematic method getting along 
with computer to construct a Rabi oscillation in the general case. 

To begin with, let us make a brief review of the model. 
We consider an atom with $n$ energy levels 
$
\{(\ket{k},E_{k})\ |\ 0 \leq k \leq n-1\}
$ 
which interacts with $n-1$ external fields. We set 
$
\Delta_{k}\equiv E_{k}-E_{0}\ \mbox{for}\ 1 \leq k \leq n-1
$ 
and assume the condition
\[
E_{1}-E_{0} > E_{2}-E_{1} > \cdots > E_{n-1}-E_{n-2}
\]
for simplicity.

We subject the atom to $n-1$ laser fields having the frequencies $\omega_{k}$ 
equal to the energy differences 
$\Delta_{k}-\Delta_{k-1}=E_{k}-E_{k-1}$. 
As an image see the following picture : 

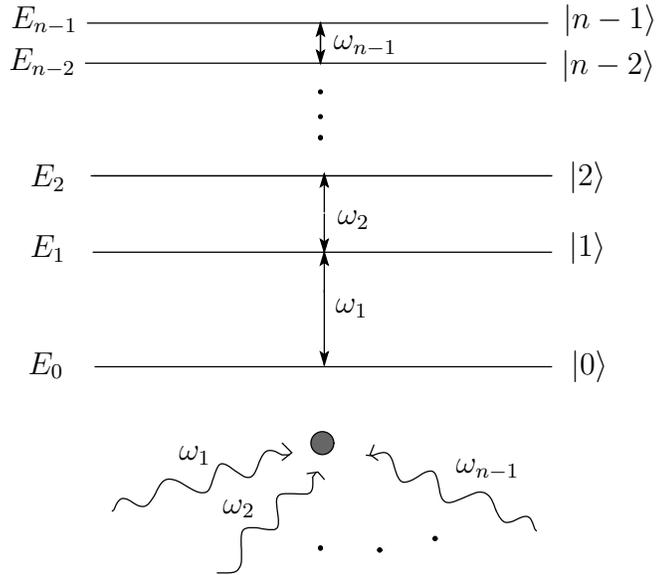
\vspace{5mm}
\begin{center}
\input{atom-with-n-energy-levels.fig}
\end{center}

\par \noindent
Then the evolution operator $U(t)$ defined by the Schr{\" o}dinger equation
\[
i\frac{d}{dt}U(t)=HU(t)\qquad (\hbar=1),
\]
where $H$ is the Hamiltonian given in \cite{FHKW2} (we don't repeat it here), 
is given by
\begin{equation}
\label{eq:evolution}
U(t)=\mbox{e}^{-itE_{0}}V^{\dagger}\mbox{e}^{-itC},
\end{equation}
where $V=V(t)$ is
\[
V=
\left(
  \begin{array}{ccccccc}
    1 &    &        &    &    &   &                                    \\
      & \mbox{e}^{i(\omega_{1}t+\phi_{1})} &  &   &   &  &             \\
      &    & \mbox{e}^{i(\omega_{1}t+\omega_{2}t+\phi_{1}+\phi_{2})} 
      &  &  &  &                                                       \\
      &    &           &  \cdot  &     &  &                            \\
      &    &           &         & \quad \quad \cdot   &    &          \\
      &    &  &  &  & \mbox{e}^{i\left(\sum_{k=1}^{n-2}\omega_{k}t+
      \sum_{k=1}^{n-2}\phi_{k}\right)} & \\
      &    &  &  &  &   &  \mbox{e}^{i\left(\sum_{k=1}^{n-1}\omega_{k}t+
      \sum_{k=1}^{n-1}\phi_{k}\right)} 
  \end{array}
\right)
\]
and the constant matrix $C$ consisting of all coupling constants
\begin{equation}
\label{eq:matrix-C}
C\equiv C(g_{1},g_{2},\cdots,g_{n-1})=
\left(
  \begin{array}{ccccccc}
    0 & g_{1} &   &   &   &   &             \\
    g_{1} & 0 & g_{2} &   &   &   &         \\
      & g_{2} & 0 & g_{3} &   &   &         \\
      &   & \cdot & \cdot & \cdot &  &      \\
      &   &   & \cdot & \cdot & \cdot &     \\
      &   &   &   & g_{n-2} & 0 & g_{n-1}   \\
      &   &   &   &   & g_{n-1} & 0
  \end{array}
\right).
\end{equation}
Therefore the remaining problem is to calculate the exponential 
$\mbox{e}^{-itC}$, which is however very hard. 

In \cite{FHKW2} we determined it by use of diagonalization method of matrices 
in the case of n = 3, 4 and 5. If all coupling constants are equal 
($g_{1}=g_{2}=\cdots=g_{n-1}$) then the situation becomes very easy, see for 
example \cite{FHKW1}. 
In the following we present a method to calculate $\mbox{e}^{-itC}$ in the 
general case, which is universal in a sense. 

\par \vspace{5mm}
Now let us review the formula in \cite{FO} within our necessity. 
Let $A$ be a matrix in $M(n,\fukuso)$ and $\{\lambda_{1},\lambda_{2},\cdots,
\lambda_{n}\}$ the set of eigenvalues (containing multiplicities) of $A$.
Then the exponential $\mbox{e}^{-itA}$ is given by the clear formula
\begin{equation}
\label{eq:last-formula}
\mbox{e}^{-itA}=f_{0}(t){\bf 1}_{n}+f_{1}(t)A+f_{2}(t)A^{2}+\cdots +
f_{n-1}(t)A^{n-1}
\end{equation}
with
\begin{equation}
\label{eq:coefficients}
f_{l}(t)=(-1)^{n+1}\sum_{k=1}^{n}
\frac{\left(p_{n-l-1}\right)_{k}\mbox{e}^{-it\lambda_{k}}}
{\prod_{j=1,j\ne k}^{n}(\lambda_{j}-\lambda_{k})} \quad \mbox{for}\quad 0\leq l\leq n-1.
\end{equation}
Here $\{p_{1},p_{2},\cdots,p_{n-1},p_{n}\}$ are a kind of fundamental 
symmetric polynomials of $\{\lambda_{1},\lambda_{2},\cdots,\lambda_{n}\}$ and 
$\{(p_{1})_{k},(p_{2})_{k},\cdots,(p_{n-1})_{k}\}$ are them consisting of 
$\{\lambda_{1},\cdots,\lambda_{k-1},\lambda_{k+1},\cdots,\lambda_{n}\}$.
For example, 

\par \noindent
\underline{n=3} :
\begin{equation}
\mbox{e}^{-itA}=f_{0}(t){\bf 1}_{3}+f_{1}(t)A+f_{2}(t)A^{2}
\end{equation}
with
\begin{eqnarray*}
f_{0}(t)&=&
\frac{\lambda_{2}\lambda_{3}\mbox{e}^{-it\lambda_{1}}}
{(\lambda_{2}-\lambda_{1})(\lambda_{3}-\lambda_{1})}
+
\frac{\lambda_{1}\lambda_{3}\mbox{e}^{-it\lambda_{2}}}
{(\lambda_{1}-\lambda_{2})(\lambda_{3}-\lambda_{2})}
+
\frac{\lambda_{1}\lambda_{2}\mbox{e}^{-it\lambda_{3}}}
{(\lambda_{1}-\lambda_{3})(\lambda_{2}-\lambda_{3})}, \\
f_{1}(t)&=&
-\frac{(\lambda_{2}+\lambda_{3})\mbox{e}^{-it\lambda_{1}}}
{(\lambda_{2}-\lambda_{1})(\lambda_{3}-\lambda_{1})}
-
\frac{(\lambda_{1}+\lambda_{3})\mbox{e}^{-it\lambda_{2}}}
{(\lambda_{1}-\lambda_{2})(\lambda_{3}-\lambda_{2})}
-
\frac{(\lambda_{1}+\lambda_{2})\mbox{e}^{-it\lambda_{3}}}
{(\lambda_{1}-\lambda_{3})(\lambda_{2}-\lambda_{3})}, \\
f_{2}(t)&=&
\frac{\mbox{e}^{-it\lambda_{1}}}
{(\lambda_{2}-\lambda_{1})(\lambda_{3}-\lambda_{1})}
+
\frac{\mbox{e}^{-it\lambda_{2}}}
{(\lambda_{1}-\lambda_{2})(\lambda_{3}-\lambda_{2})}
+
\frac{\mbox{e}^{-it\lambda_{3}}}
{(\lambda_{1}-\lambda_{3})(\lambda_{2}-\lambda_{3})}.
\end{eqnarray*}

\par \vspace{5mm} \noindent
\underline{n=4} :
\begin{equation}
\mbox{e}^{-itA}=f_{0}(t){\bf 1}_{4}+f_{1}(t)A+f_{2}(t)A^{2}+f_{3}(t)A^{3}
\end{equation}
with
\begin{eqnarray*}
f_{0}(t)&=&
\frac{\lambda_{2}\lambda_{3}\lambda_{4}\mbox{e}^{-it\lambda_{1}}}
{(\lambda_{2}-\lambda_{1})(\lambda_{3}-\lambda_{1})(\lambda_{4}-\lambda_{1})}
+
\frac{\lambda_{1}\lambda_{3}\lambda_{4}\mbox{e}^{-it\lambda_{2}}}
{(\lambda_{1}-\lambda_{2})(\lambda_{3}-\lambda_{2})(\lambda_{4}-\lambda_{2})}
\nonumber \\
&&+
\frac{\lambda_{1}\lambda_{2}\lambda_{4}\mbox{e}^{-it\lambda_{3}}}
{(\lambda_{1}-\lambda_{3})(\lambda_{2}-\lambda_{3})(\lambda_{4}-\lambda_{3})}
+
\frac{\lambda_{1}\lambda_{2}\lambda_{3}\mbox{e}^{-it\lambda_{4}}}
{(\lambda_{1}-\lambda_{4})(\lambda_{2}-\lambda_{4})(\lambda_{3}-\lambda_{4})}, 
\\
f_{1}(t)&=&
-\frac{(\lambda_{2}\lambda_{3}+\lambda_{2}\lambda_{4}+\lambda_{3}\lambda_{4})
\mbox{e}^{-it\lambda_{1}}}
{(\lambda_{2}-\lambda_{1})(\lambda_{3}-\lambda_{1})(\lambda_{4}-\lambda_{1})}
-
\frac{(\lambda_{1}\lambda_{3}+\lambda_{1}\lambda_{4}+\lambda_{3}\lambda_{4})
\mbox{e}^{-it\lambda_{2}}}
{(\lambda_{1}-\lambda_{2})(\lambda_{3}-\lambda_{2})(\lambda_{4}-\lambda_{2})}
\nonumber \\
&&-
\frac{(\lambda_{1}\lambda_{2}+\lambda_{1}\lambda_{4}+\lambda_{2}\lambda_{4})
\mbox{e}^{-it\lambda_{3}}}
{(\lambda_{1}-\lambda_{3})(\lambda_{2}-\lambda_{3})(\lambda_{4}-\lambda_{3})}
-
\frac{(\lambda_{1}\lambda_{2}+\lambda_{1}\lambda_{3}+\lambda_{2}\lambda_{3})
\mbox{e}^{-it\lambda_{4}}}
{(\lambda_{1}-\lambda_{4})(\lambda_{2}-\lambda_{4})(\lambda_{3}-\lambda_{4})}, 
\\
f_{2}(t)&=&
\frac{(\lambda_{2}+\lambda_{3}+\lambda_{4})\mbox{e}^{-it\lambda_{1}}}
{(\lambda_{2}-\lambda_{1})(\lambda_{3}-\lambda_{1})(\lambda_{4}-\lambda_{1})}
+
\frac{(\lambda_{1}+\lambda_{3}+\lambda_{4})\mbox{e}^{-it\lambda_{2}}}
{(\lambda_{1}-\lambda_{2})(\lambda_{3}-\lambda_{2})(\lambda_{4}-\lambda_{2})}
\nonumber \\
&&+
\frac{(\lambda_{1}+\lambda_{2}+\lambda_{4})\mbox{e}^{-it\lambda_{3}}}
{(\lambda_{1}-\lambda_{3})(\lambda_{2}-\lambda_{3})(\lambda_{4}-\lambda_{3})}
+
\frac{(\lambda_{1}+\lambda_{2}+\lambda_{3})\mbox{e}^{-it\lambda_{4}}}
{(\lambda_{1}-\lambda_{4})(\lambda_{2}-\lambda_{4})(\lambda_{3}-\lambda_{4})},
\\
f_{3}(t)&=&
-\frac{\mbox{e}^{-it\lambda_{1}}}
{(\lambda_{2}-\lambda_{1})(\lambda_{3}-\lambda_{1})(\lambda_{4}-\lambda_{1})}
-
\frac{\mbox{e}^{-it\lambda_{2}}}
{(\lambda_{1}-\lambda_{2})(\lambda_{3}-\lambda_{2})(\lambda_{4}-\lambda_{2})}
\nonumber \\
&&-
\frac{\mbox{e}^{-it\lambda_{3}}}
{(\lambda_{1}-\lambda_{3})(\lambda_{2}-\lambda_{3})(\lambda_{4}-\lambda_{3})}
-
\frac{\mbox{e}^{-it\lambda_{4}}}
{(\lambda_{1}-\lambda_{4})(\lambda_{2}-\lambda_{4})(\lambda_{3}-\lambda_{4})}.
\end{eqnarray*}

\par \vspace{3mm}
It is notable that our formula (\ref{eq:last-formula}) with 
(\ref{eq:coefficients}) is convenient to use computer if the eigenvalues are 
known.

\par \vspace{5mm}
Next, what we do is to look for the eigenvalues of $C$ in (\ref{eq:matrix-C}) 
in order to use the formula above. The characteristic polynomial of $C$ is
\begin{equation}
\label{eq:characteristic polynomial}
f_{n}(\lambda)
\equiv |\lambda {\bf 1}_{n}-C|
=\left|
  \begin{array}{ccccccc}
    \lambda & -g_{1} &   &   &   &   &             \\
    -g_{1} & \lambda & -g_{2} &   &   &   &        \\
      & -g_{2} & \lambda & -g_{3} &   &   &        \\
      &   & \cdot & \cdot & \cdot &  &             \\
      &   &   & \cdot & \cdot & \cdot &            \\
      &   &   &   & -g_{n-2} & \lambda & -g_{n-1}  \\
      &   &   &   &   & -g_{n-1} & \lambda
  \end{array}
\right|.
\end{equation}
By using the Laplace expansion of determinant it is easy to see 
\begin{equation}
f_{n}(\lambda)=\lambda f_{n-1}(\lambda)-g_{n-1}^{2}f_{n-2}(\lambda);\quad 
f_{0}(\lambda)=1,\ f_{1}(\lambda)=\lambda.
\end{equation}
For example, 
\begin{eqnarray}
\label{eq:examples}
f_{2}(\lambda)&=&\lambda^{2}-g_{1}^{2},\nonumber \\
f_{3}(\lambda)&=&\lambda\left(\lambda^{2}-g_{1}^{2}-g_{2}^{2}\right),\nonumber \\
f_{4}(\lambda)&=&\lambda^{4}-(g_{1}^{2}+g_{2}^{2}+g_{3}^{2})\lambda^{2}+
g_{1}^{2}g_{3}^{2},\nonumber \\
f_{5}(\lambda)&=&\lambda\left(\lambda^{4}-(g_{1}^{2}+g_{2}^{2}+g_{3}^{2}+g_{4}^{2})
\lambda^{2}+(g_{1}^{2}g_{3}^{2}+g_{1}^{2}g_{4}^{2}+g_{2}^{2}g_{4}^{2})\right),
\nonumber \\
f_{6}(\lambda)&=&\lambda^{6}-(g_{1}^{2}+g_{2}^{2}+g_{3}^{2}+g_{4}^{2}+g_{5}^{2})
\lambda^{4}+(g_{1}^{2}g_{3}^{2}+g_{1}^{2}g_{4}^{2}+g_{1}^{2}g_{5}^{2}+
g_{2}^{2}g_{4}^{2}+g_{2}^{2}g_{5}^{2}+g_{3}^{2}g_{5}^{2})\lambda^{2} \nonumber \\
&&-g_{1}^{2}g_{3}^{2}g_{5}^{2} \nonumber \\
f_{7}(\lambda)&=&\lambda 
\left(
\lambda^{6}-(g_{1}^{2}+g_{2}^{2}+g_{3}^{2}+g_{4}^{2}+
g_{5}^{2}+g_{6}^{2})\lambda^{4}+(g_{1}^{2}g_{3}^{2}+g_{1}^{2}g_{4}^{2}+ 
g_{1}^{2}g_{5}^{2}+g_{1}^{2}g_{6}^{2}+g_{2}^{2}g_{4}^{2}+ 
\right. \nonumber \\
&&
\left.
g_{2}^{2}g_{5}^{2}+g_{2}^{2}g_{6}^{6}+g_{3}^{2}g_{5}^{2}+g_{3}^{2}g_{6}^{2}+
g_{4}^{2}g_{6}^{2})\lambda^{2}-(g_{1}^{2}g_{3}^{2}g_{5}^{2}+g_{1}^{2}g_{3}^{2}g_{6}^{2}+
g_{1}^{2}g_{4}^{2}g_{6}^{2}+g_{2}^{2}g_{4}^{2}g_{6}^{2})
\right). \nonumber \\
&{}&
\end{eqnarray}

Now, let us look for the general form of the characteristic polynomial of 
$f_{n}(\lambda)$.

\par \vspace{5mm} \noindent
{\bf Main Result} 
\par \noindent
(i)\ $n=2m$ :
\begin{equation}
f_{2m}(\lambda)=
\lambda^{2m}-\phi_{2}\lambda^{2m-2}+\cdots+(-1)^{k}\phi_{2k}\lambda^{2m-2k}+\cdots+
(-1)^{m-1}\phi_{2m-2}\lambda^{2}+(-1)^{m}\phi_{2m}
\end{equation}
with
\begin{eqnarray}
\phi_{2}&=&\sum_{i=1}^{2m-1}g_{i}^{2}, \nonumber \\
\phi_{2k}&=&\sum_{1\leq i_{1}<i_{2}<i_{3}<\cdots <i_{k-1}<i_{k};\ |i_{1}-i_{2}|\geq 2,
|i_{2}-i_{3}|\geq 2,\cdots,|i_{k-1}-i_{k}|\geq 2}^{2m-1} 
g_{i_{1}}^{2}g_{i_{2}}^{2}\cdots g_{i_{k-1}}^{2}g_{i_{k}}^{2} \nonumber \\ 
&{}&\mbox{for}\quad 2\leq k\leq m-1.
\end{eqnarray}
\par \vspace{5mm} \noindent
(ii)\ $n=2m+1$ :
\begin{equation}
f_{2m+1}(\lambda)=\lambda
\left\{
\lambda^{2m}-\varphi_{2}\lambda^{2m-2}+\cdots+(-1)^{k}\varphi_{2k}\lambda^{2m-2k}+
\cdots+(-1)^{m-1}\varphi_{2m-2}\lambda^{2}+(-1)^{m}\varphi_{2m}
\right\}
\end{equation}
with
\begin{eqnarray}
\varphi_{2}&=&\sum_{i=1}^{2m}g_{i}^{2}, \nonumber \\
\varphi_{2k}&=&\sum_{1\leq i_{1}<i_{2}<i_{3}<\cdots <i_{k-1}<i_{k};\ |i_{1}-i_{2}|\geq 2,
|i_{2}-i_{3}|\geq 2,\cdots,|i_{k-1}-i_{k}|\geq 2}^{2m} 
g_{i_{1}}^{2}g_{i_{2}}^{2}\cdots g_{i_{k-1}}^{2}g_{i_{k}}^{2} \nonumber \\
&{}&\mbox{for}\quad 2\leq k\leq m.
\end{eqnarray}

\par \vspace{3mm}
The key point of the proof is the following equations. 
\begin{eqnarray*}
\phi_{2k}
&=&\sum_{1\leq i_{1}<i_{2}<i_{3}<\cdots <i_{k-1}<i_{k};\ |i_{1}-i_{2}|\geq 2,
|i_{2}-i_{3}|\geq 2,\cdots,|i_{k-1}-i_{k}|\geq 2}^{2m-1} 
g_{i_{1}}^{2}g_{i_{2}}^{2}\cdots g_{i_{k-1}}^{2}g_{i_{k}}^{2} \\
&=&\sum_{1\leq i_{1}<i_{2}<i_{3}<\cdots <i_{k-1};\ |i_{1}-i_{2}|\geq 2,
|i_{2}-i_{3}|\geq 2,\cdots,|i_{k-2}-i_{k-1}|\geq 2}^{2m-3} 
g_{i_{1}}^{2}g_{i_{2}}^{2}\cdots g_{i_{k-1}}^{2}g_{2m-1}^{2}+ \\
&{}& \sum_{1\leq i_{1}<i_{2}<i_{3}<\cdots <i_{k-1}<i_{k};\ |i_{1}-i_{2}|\geq 2,
|i_{2}-i_{3}|\geq 2,\cdots,|i_{k-1}-i_{k}|\geq 2}^{2m-2} 
g_{i_{1}}^{2}g_{i_{2}}^{2}\cdots g_{i_{k-1}}^{2}g_{i_{k}}^{2} 
\end{eqnarray*}
and
\begin{eqnarray*}
\varphi_{2k}
&=&\sum_{1\leq i_{1}<i_{2}<i_{3}<\cdots <i_{k-1}<i_{k};\ |i_{1}-i_{2}|\geq 2,
|i_{2}-i_{3}|\geq 2,\cdots,|i_{k-1}-i_{k}|\geq 2}^{2m} 
g_{i_{1}}^{2}g_{i_{2}}^{2}\cdots g_{i_{k-1}}^{2}g_{i_{k}}^{2} \\
&=&\sum_{1\leq i_{1}<i_{2}<i_{3}<\cdots <i_{k-1};\ |i_{1}-i_{2}|\geq 2,
|i_{2}-i_{3}|\geq 2,\cdots,|i_{k-2}-i_{k-1}|\geq 2}^{2m-2} 
g_{i_{1}}^{2}g_{i_{2}}^{2}\cdots g_{i_{k-1}}^{2}g_{2m}^{2}+ \\
&{}& \sum_{1\leq i_{1}<i_{2}<i_{3}<\cdots <i_{k-1}<i_{k};\ |i_{1}-i_{2}|\geq 2,
|i_{2}-i_{3}|\geq 2,\cdots,|i_{k-1}-i_{k}|\geq 2}^{2m-1} 
g_{i_{1}}^{2}g_{i_{2}}^{2}\cdots g_{i_{k-1}}^{2}g_{i_{k}}^{2}.
\end{eqnarray*}
For the proof we have only to use the mathematical induction. We leave 
the remaining part to readers. 

\par \vspace{3mm}
A comment is in order. Since the matrix $C$ in (\ref{eq:matrix-C}) is 
real symmetric (of course hermitian) its eigenvalues are all {\bf real}.

\par \vspace{5mm}
Next, let us solve the characteristic polynomial of $C$. From (\ref{eq:examples}) 
\par \noindent
\underline{$n=2$} :
\[
\lambda_{1}=g_{1},\quad \lambda_{2}=-g_{1}.
\]
\par \noindent
\underline{$n=3$} :
\[
\lambda_{1}=\sqrt{g_{1}^{2}+g_{2}^{2}},\quad \lambda_{2}=0,
\quad \lambda_{3}=-\sqrt{g_{1}^{2}+g_{2}^{2}}.
\]
\par \noindent
\underline{$n=4$} :
\[
\lambda_{1}=\frac{\sqrt{A}+\sqrt{B}}{2},\quad 
\lambda_{2}=\frac{\sqrt{A}-\sqrt{B}}{2},\quad 
\lambda_{3}=-\frac{\sqrt{A}-\sqrt{B}}{2},\quad 
\lambda_{4}=-\frac{\sqrt{A}+\sqrt{B}}{2}
\]
where
\[
A=g_{2}^{2}+(g_{1}+g_{3})^{2},\quad B=g_{2}^{2}+(g_{1}-g_{3})^{2}.
\]
\par \noindent
\underline{$n=5$} :
\[
\lambda_{1}=\frac{\sqrt{A}+\sqrt{B}}{2},\quad 
\lambda_{2}=\frac{\sqrt{A}-\sqrt{B}}{2},\quad 
\lambda_{3}=0,\quad
\lambda_{4}=-\frac{\sqrt{A}-\sqrt{B}}{2},\quad 
\lambda_{5}=-\frac{\sqrt{A}+\sqrt{B}}{2}
\]
where
\[
A=g_{1}^{2}+g_{2}^{2}+g_{3}^{2}+g_{4}^{2}+2\sqrt{g_{1}^{2}g_{3}^{2}+
g_{1}^{2}g_{4}^{2}+g_{2}^{2}g_{4}^{2}},\quad 
B=g_{1}^{2}+g_{2}^{2}+g_{3}^{2}+g_{4}^{2}-2\sqrt{g_{1}^{2}g_{3}^{2}+
g_{1}^{2}g_{4}^{2}+g_{2}^{2}g_{4}^{2}}.
\]
\par \noindent
\underline{$n=6$} :

By setting $\lambda^{2}=x$ we can use the Cardano formula (see for example 
\cite{KF5}). The solutions of the equation
\[
x^{3}-ax^{2}+bx-c=0 
\]
with
\[
a=g_{1}^{2}+g_{2}^{2}+g_{3}^{2}+g_{4}^{2}+g_{5}^{2},\
b=g_{1}^{2}g_{3}^{2}+g_{1}^{2}g_{4}^{2}+g_{1}^{2}g_{5}^{2}+
g_{2}^{2}g_{4}^{2}+g_{2}^{2}g_{5}^{2}+g_{3}^{2}g_{5}^{2},\
c=g_{1}^{2}g_{3}^{2}g_{5}^{2}
\]
are given by
\[
x_{1}=u_{0}+v_{0}+\frac{a}{3},\quad
x_{2}=\sigma u_{0}+\sigma^{2}v_{0}+\frac{a}{3},\quad
x_{3}=\sigma^{2}u_{0}+\sigma v_{0}+\frac{a}{3},
\]
where $\sigma=\mbox{e}^{2\pi i/3}$ and $u_{0}$, $v_{0}$ are each solution 
of the binomial equations
\[
u^{3}=\frac{-q+\sqrt{q^{2}+4p^{3}}}{2},\quad 
v^{3}=\frac{-q-\sqrt{q^{2}+4p^{3}}}{2}
\]
with
\[
p=\frac{b}{3}-\frac{a^{2}}{9},\quad
q=-c+\frac{ab}{3}-\frac{2a^{3}}{27}.
\]
It is not difficult to show that we can choose $x_{1}\geq x_{2}\geq x_{3} >0$. 
Therefore the solutions are
\[
\lambda_{1}=\sqrt{x_{1}},\ \lambda_{2}=\sqrt{x_{2}},\ 
\lambda_{3}=\sqrt{x_{3}},\ 
\lambda_{4}=-\sqrt{x_{3}},\ \lambda_{5}=-\sqrt{x_{2}},\ 
\lambda_{6}=-\sqrt{x_{1}}.
\]
\par \noindent
\underline{$n=7$} :

In the case of $n=6$ we have only to change $a$, $b$ and $c$ to
\begin{eqnarray*}
a&=&g_{1}^{2}+g_{2}^{2}+g_{3}^{2}+g_{4}^{2}+g_{5}^{2}+g_{6}^{2},\\
b&=&
g_{1}^{2}g_{3}^{2}+g_{1}^{2}g_{4}^{2}+g_{1}^{2}g_{5}^{2}+g_{1}^{2}g_{6}^{2}+
g_{2}^{2}g_{4}^{2}+g_{2}^{2}g_{5}^{2}+g_{2}^{2}g_{6}^{2}+
g_{3}^{2}g_{5}^{2}+g_{3}^{2}g_{6}^{2}+g_{4}^{2}g_{6}^{2},\\
c&=&g_{1}^{2}g_{3}^{2}g_{5}^{2}+g_{1}^{2}g_{3}^{2}g_{6}^{2}+
g_{1}^{2}g_{4}^{2}g_{6}^{2}+g_{2}^{2}g_{4}^{2}g_{6}^{2}
\end{eqnarray*}
and obtain
\[
\lambda_{1}=\sqrt{x_{1}},\ \lambda_{2}=\sqrt{x_{2}},\ 
\lambda_{3}=\sqrt{x_{3}},\ 
\lambda_{4}=0,\
\lambda_{5}=-\sqrt{x_{3}},\ \lambda_{6}=-\sqrt{x_{2}},\ 
\lambda_{7}=-\sqrt{x_{1}}.
\]

\par \vspace{5mm}
For $n=8$ and $9$ we can in principle solve the characteristic polynomial 
(\ref{eq:characteristic polynomial}) by use of the Ferrari or Euler formula 
(see \cite{KF5}), which is of course very complicated. 
However, for $n\geq 10$ it is impossible to obtain algebraic solutions of the
characteristic polynomial by the famous Galois theory. 
Therefore we must appeal to some approximation method like the Newton's one 
(which is well--known), see the following picture.

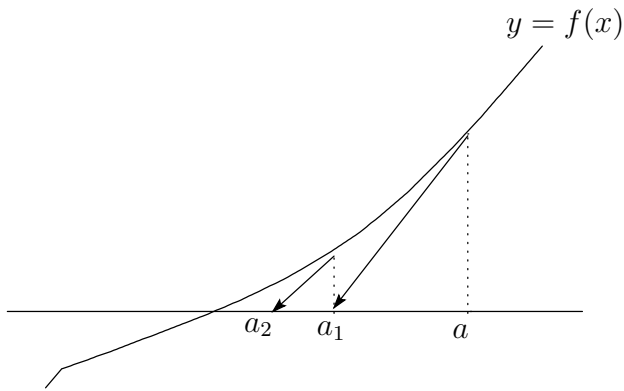
\vspace{5mm}
\begin{center}
\input{newton-approximation-method.fig}
\end{center}

\par \noindent
Here for $a$ given, the points $a_{1}$, $a_{2}$, etc are given by
\[
a_{1}=a-\frac{f(a)}{f^{'}(a)},\ 
a_{2}=a_{1}-\frac{f(a_{1})}{f^{'}(a_{1})},\ \cdots\ ,\
a_{k}=a_{k-1}-\frac{f(a_{k-1})}{f^{'}(a_{k-1})},\ \cdots\ .
\]
For an appropriate number $n$ we have only to set $\lambda=a_{n}$. 
By changing $a$ in the general case we obtain a set of approximate solutions 
$\{\lambda_{1},\lambda_{2},\cdots,\lambda_{n}\}$.

\par \vspace{10mm}
In this paper we generalized the result in \cite{FHKW2} by making use of 
\cite{FO}. If we can find exact or approximate solutions of 
the characteristic polynomial (\ref{eq:characteristic polynomial}) then 
we have the (exact or approximate) evolution operator by 
(\ref{eq:last-formula}) and (\ref{eq:coefficients}). 
That is, this means that a complicated unitary matrix in qudit theory was 
obtained. 
It may be possible to replace long quantum logic gates in qudit theory 
with few unitary matrices constructed in the paper, which will be discussed 
in another paper.

%%%%%%%%%%%%%
%References%
%%%%%%%%%%%%%

\end{document}